\documentclass{webofc}
\usepackage[varg]{txfonts}   
\usepackage{lipsum}  
\usepackage{graphicx}
\usepackage{amsmath}
\usepackage{amssymb}
\usepackage{natbib}
\DeclareFontFamily{OT1}{pzc}{}
\DeclareFontShape{OT1}{pzc}{m}{it}{ <-> s*[1.15] pzcmi7t }{}
\DeclareMathAlphabet{\mathpzc}{OT1}{pzc}{m}{it}

\newcommand{\wbar}{{\overline \tau}}
\newcommand{\Mbar}{{\overline{\mathcal M}}\hspace{0.5mm}} 
\begin{document}
\setlength{\footskip}{3.60004pt}.
\title{Far-from-equilibrium attractors for massive kinetic theory in the relaxation time approximation}
%
%

\author{\firstname{Huda} \lastname{Alalawi}\inst{1}\fnsep\thanks{\email{halalawi@kent.edu}} \and
        \firstname{Michael} \lastname{Strickland}\inst{2}\fnsep\thanks{\email{mstrick6@kent.edu}} 
}

\institute{Department of Physics, Kent State University, Kent, Ohio 44242, USA}

\abstract{%
In this proceedings contribution, we summarize recent findings concerning the presence of early- and late-time attractors in non-conformal kinetic theory. We study the effects of varying both the initial momentum-space anisotropy and initialization times using an exact solution of the 0+1D boost-invariant Boltzmann equation with a mass- and temperature-dependent relaxation time. Our findings support the existence of a longitudinal pressure attractor, but they do not support the existence of distinct attractors for the bulk viscous and shear pressures. Considering a large set of integral moments, we show that for moments with greater than one power of longitudinal momentum squared, both early- and late-time attractors are present.}
\maketitle
\section{Introduction}
\label{intro}

A fundamental question in far-from-equilibrium relativistic dynamics in recent decades has been the extent to which relativistic viscous hydrodynamics can represent such dynamics. Exact solutions within the relaxation time approximation (RTA)  to the relativistic Boltzmann equation have proven critical in evaluating the accuracy of dissipative hydrodynamic models. These solutions also shed light on the evolution of far-from-equilibrium attractors, which merge smoothly onto viscous hydrodynamics at late times, yet remain valid earlier when linearized treatments fail.

While most studies have concentrated on conformal systems, there is growing interest in non-conformal systems where the dynamics are influenced by more than one scale. In this proceedings contribution, we extend prior exact solutions for a massive gas~\cite{Florkowski:2014sfa} by computing all moments of the one-particle distribution function.  In addition,  we use a temperature- and mass-dependent relaxation time and search for separate attractors for the shear and bulk viscous corrections.  The work we report on, which originally appeared in Ref.~\cite{Alalawi:2022pmg}, goes beyond prior works in which it was assumed that the relaxation time was either constant or inversely proportional to the temperature.

We found in Ref.~\cite{Alalawi:2022pmg} that moments with greater than one power of the longitudinal momentum squared, possess both forward and early-time (pull-back) attractors. However, we found that although the shear and bulk viscous corrections do not have early-time attractors on their own, the difference between them does, indicating an attractor in the scaled longitudinal pressure. These findings are consistent with and expand upon findings of other groups.

\section{Setup}
\label{sec-1}

We will assume Bjorken flow, in which case in Milne coordinates one has $u^\tau=1$ and $u^{x,y,\varsigma}=0$, where $\tau $ is the longitudinal proper-time, $\tau = \sqrt{t^2-z^2}$, and $\varsigma$ is the spatial rapidity, \mbox{$\varsigma = \tanh^{-1}(z/t)$}. Therefore, all scalar quantities depend only on the longitudinal proper time $\tau$. We start with the RTA Boltzmann equation 
\begin{equation}
\centering
 p^\mu \partial_\mu  f(x,p) = \frac{p \cdot u}{\tau_{\rm eq}} \left( f_{\rm eq}-f \right)=C[f] \, , 
\label{kineq}
\end{equation}
where $p^\mu$ is the particle four-momentum, $u^\mu$ is the four-velocity of the local rest frame, \mbox{$a \cdot b \equiv a^\mu b_\mu$}, and $C$ is the collisional kernel. This equation takes a simple form when it is written in terms of boost-invariant variables~\cite{Florkowski:2013lza,Florkowski:2014sfa}
\begin{equation}
\centering
\frac{\partial f}{\partial \tau}  = 
\frac{f_{\rm eq}-f}{\tau_{\rm eq}} \, .
\label{eq:simpleformrta}
\end{equation} 
Here, the exact solution to eq.~\eqref{eq:simpleformrta} is~\cite{Florkowski:2013lza}
\begin{equation}
f(\tau,w,p_T) = D(\tau,\tau_0) f_0(w,p_T)  + \int_{\tau_0}^\tau \frac{d\tau^\prime}{\tau_{\rm eq}(\tau^\prime)} \, D(\tau,\tau^\prime) \, 
f_{\rm eq}(\tau^\prime,w,p_T) \, ,  \label{eq:solf}
\end{equation}
where the damping function $D$ is defined as $D(\tau_2,\tau_1) = \exp\left[-\int_{\tau_1}^{\tau_2}
\frac{d\tau^{\prime\prime}}{\tau_{\rm eq}(\tau^{\prime\prime})} \right]$, $f$ is the one-particle distribution function and $f_{\rm eq}$ is the equilibrium distribution that is assumed to be a Boltzmann distribution.  When written in terms of the boost-invariant variables, the equilibrium distribution function is
\begin{equation}
\centering
f_{\rm eq}(\tau,w,p_T) =
\exp\!\left[
- \frac{\sqrt{w^2+ (p_T^2+m^2)\tau^2}}{T \tau}  \right].
\label{eqdistform}
\end{equation}
The initial distribution function $f_0(w,p_T)$ is specified at $\tau = \tau_0$ and assumed to be of spheroidally-deformed form~\cite{Romatschke:2003ms,Romatschke:2004jh}
\begin{equation}
    f_0(w,p_T) = 
\exp\left[
-\frac{\sqrt{(p\cdot u)^2 + \xi_0 (p\cdot z)^2}}{\Lambda_0} \, \right] \nonumber 
= 
\exp\left[
-\frac{\sqrt{(1+\xi_0) w^2 + (m^2+p_T^2) \tau_0^2}}{\Lambda_0 \tau_0}\, \right] ,
\label{G0}
\end{equation}
where $\xi_0$ is the initial anisotropy parameter and $\Lambda_0$ is the initial transverse momentum scale. The relaxation time $\tau_{\rm eq}$ is defined as $\tau_{\rm eq}(T,m)=5\bar{\eta}\,\gamma(\hat{m})/T$ with $\gamma(\hat{m}) \equiv \frac{3}{\kappa(\hat{m})} \bigg(1+\frac{\varepsilon}{P}\bigg).$
By plotting $\gamma(\hat{m})$, we observed that $\gamma(\hat{m})$ goes to unity in the massless limit and grows linearly at large $m/T$, which corresponds either to fixed temperature and large mass or fixed mass and small temperature~\cite{Alalawi:2022pmg}. We note that $\gamma(\hat{m}) \geq 1$ implies that a massive gas always relaxes more slowly to equilibrium than a massless one.

To understand the properties and dynamics of the system, we will work with the general moments of the one-particle distribution function~\cite{Strickland:2018ayk,Alalawi:2022pmg,Alalawi:2020zbx}, which can be expressed in terms of the boost-invariant variables as
\begin{equation}
{\cal M}^{nl}[f] = \frac{1}{(2\pi)^3 \, \tau^{n+2l}} \int  dw \, d^2p_T  \, v^{n-1} w^{2l} \, f(\tau,w,p_T)\, ,
\end{equation}
and these general moments scaled by their equilibrium values, i.e., ${\Mbar}^{nl} \equiv {\cal M}^{nl}/{\cal M}^{nl}_{\rm eq}$ where in the late-time limit ($\tau \rightarrow \infty$), if the system approaches equilibrium, then ${\Mbar}^{nl} \rightarrow 1$. By taking a general moment of eq.~(\ref{eq:solf}) and evaluating the integrals necessary, one obtains
\begin{eqnarray}
{\cal M}^{nl} &=&  \frac{D(\tau,\tau_0) \Lambda_0^{n+2l+2}}{(2\pi)^2}\tilde{H}^{nl}\left(\frac{\tau_0}{\tau \sqrt{1+\xi_0}}, \frac{m}{\Lambda_0} \right) \nonumber \\
&& \hspace{2cm} + \frac{1}{(2\pi)^2} \int_{\tau_0}^\tau \frac{d\tau^\prime}{\tau_{\rm eq}(\tau^\prime)} \, D(\tau,\tau^\prime) \, 
T^{n+2l+2}(\tau^\prime) \, \tilde{H}^{nl}\left(\frac{\tau'}{\tau}, \frac{m}{T(\tau^\prime)} \right)  \, . \label{eq:mnleq}
\end{eqnarray}
Finally, specializing to the case $n=2$ and $l=0$ and enforcing Landau matching \mbox{$\varepsilon(\tau) =\varepsilon_{\rm eq}(T)$}, one obtains the following integral equation
\begin{eqnarray}
&& \hspace{-1cm} 2 T^4(\tau) \, \hat{m}^2
 \left[ 3 K_2\!\left( \frac{m}{T(\tau)} \right) + \hat{m} K_1\!\left( \frac{m}{T(\tau)} \right) \right]
 \nonumber \\ 
 && = D(\tau,\tau_0) \Lambda_0^4 \tilde{H}^{20}\!\left(\frac{\tau_0}{\tau\sqrt{1{+}\xi_0}},\frac{m}{\Lambda_0}\right) +  \int_{\tau_0}^\tau \frac{d\tau^\prime}{\tau_{\rm eq}(\tau^\prime)} \, D(\tau,\tau^\prime) \, 
T^4(\tau^\prime) \tilde{H}^{20}\!\left(\frac{\tau'}{\tau}, \frac{m}{T(\tau^\prime)} \right) .
\label{eq:20final}
\end{eqnarray}
Since we obtain all the scaled moments, we can compute the viscous corrections and scaling by the equilibrium pressure. One finds that the bulk viscous correction can be expressed as $\tilde\Pi \equiv \Pi/P = -m^2 \left({\cal M}^{00} - {\cal M}^{00}_{\rm eq}\right)/\left({\cal M}^{20}_{\rm eq} - m^2 {\cal M}^{00}_{\rm eq}\right)$ and the shear viscous correction as $\tilde\pi \equiv \pi/P = 1 - \Mbar^{01} + \tilde\Pi$.
Next, we obtain the following expression for the scaled moments in the Navier-Stokes limit by adding the shear and bulk corrections to the equilibrium result and scaling by the equilibrium moments within the 14-moment approximation~\cite{grad_1949}
\begin{eqnarray}
\Mbar^{nl,\rm NS}_{\rm 14-moment} &=& 1 + \frac{1}{15 \, \wbar \, T^2 \, \gamma(\hat{m})}\frac{\left[  {\cal M}_{\rm eq}^{n+2,l} - 3 {\cal M}_{\rm eq}^{n,l+1} - m^2 {\cal M}_{\rm eq}^{n,l} \right]}{{\cal M}_{\rm eq}^{n,l}} \nonumber \\
&& \hspace{3cm} + \frac{1}{3 \, \wbar \, T } \frac{\left[ m^2 
{\cal M}_{\rm eq}^{n-1,l} - \left(1-3c_s^2\right) {\cal M}_{\rm eq}^{n+1,l} \right]}{{\cal M}_{\rm eq}^{n,l}} \, , \;\;\;
\label{eq:14momfinal}
\end{eqnarray}
and Chapman-Enskog approximation~\cite{Chapman1991-qu}
\begin{eqnarray}
\Mbar^{nl,\rm NS}_{\rm CE} &=& 1 + \frac{\varepsilon + P}{15 \, \wbar \, \gamma(\hat{m}) \, I_{42}^{(1)} } \frac{\left[  {\cal M}_{\rm eq}^{n+1,l} - 3 {\cal M}_{\rm eq}^{n-1,l+1} - m^2 {\cal M}_{\rm eq}^{n-1,l} \right]}{{\cal M}_{\rm eq}^{n,l}} \nonumber \\
&& \hspace{4cm} + \frac{1}{3 \, \wbar \, T } \frac{\left[ m^2 
{\cal M}_{\rm eq}^{n-1,l} - \left(1-3c_s^2\right) {\cal M}_{\rm eq}^{n+1,l} \right]}{{\cal M}_{\rm eq}^{n,l}} \, . \;\;\;
\label{eq:cefinal}
\end{eqnarray}

\begin{figure}
    \centering
        \includegraphics[width=1\linewidth]{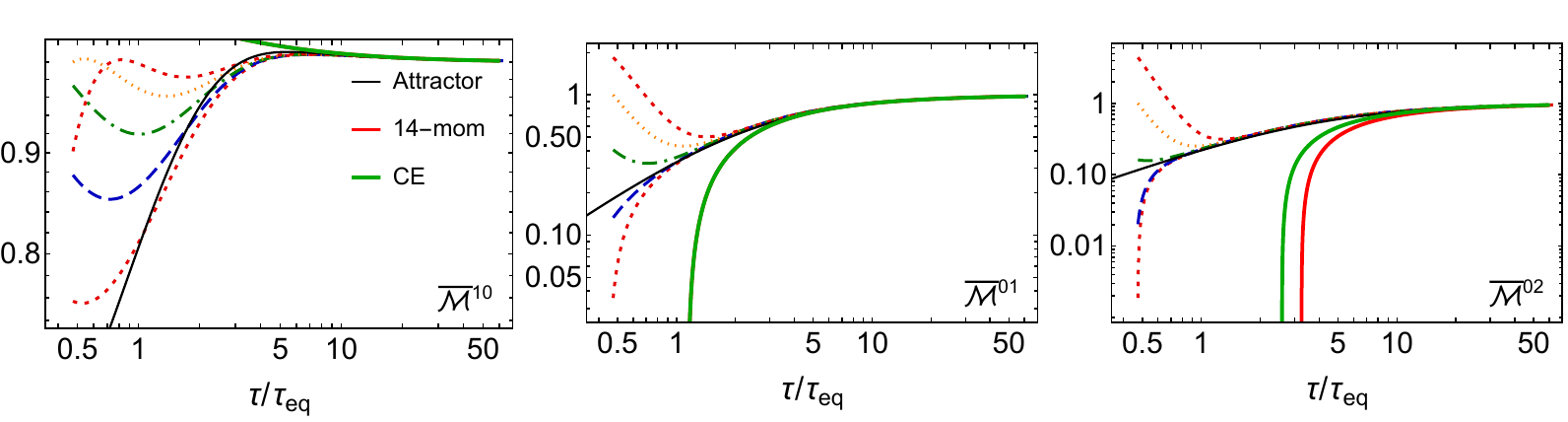}
    \includegraphics[width=1\linewidth]{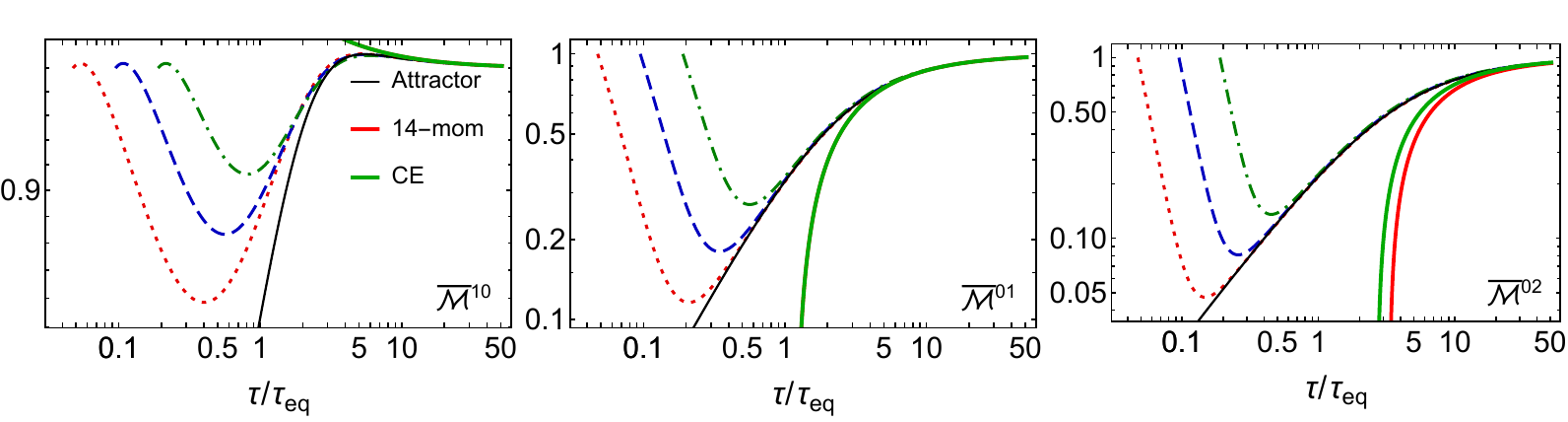}
    \caption{Scaled moments $\Mbar^{nl}$ as a function of rescaled time for the case $m = 1$~GeV and $T_0 = 1$~GeV. The solid black lines correspond to the attractor solution, the solid red lines are the first-order 14-moment predictions in eq.~\eqref{eq:14momfinal}, and the solid green lines are the first-order Chapman-Enskog predictions in eq.~\eqref{eq:cefinal}. Top: The non-solid lines are specific initial conditions initialized at $\tau_0 = 0.1$~fm/c with  $\alpha_0 = 1/\sqrt{1+\xi_0} \in \{0.12,0.25,0.5,1,2\}$. Bottom: the non-solid lines are specific initial conditions initialized with $\xi_0=0$ at $\tau_0 \in \{0.01,0.02,0.04\}$~fm/c.  Note that for $\Mbar^{10}$ and $\Mbar^{01}$, the 14-moment and Chapman-Enskog predictions are the precisely the same.}
    \label{fig:enter-label}
\end{figure}

\section{Results and Conclusions}
\label{sect:conclusions}

In this proceedings contribution, we report on our extension of prior research into the presence of attractors in non-conformal kinetic theory. We explored the time evolution of a large set of integral moments of the distribution function using an explicit solution to the boost-invariant Boltzmann equation within the RTA and released the computational code along with the paper~\cite{Alalawi:2022pmg}. Our findings, summarized in fig.~\ref{fig:enter-label}, are consistent with recent research that found late- and early-time attractors in the scaled longitudinal pressure but not in the shear and bulk viscous corrections when taken separately.

In contrast to the conformal situation, our results indicate that no early-time attractor exists for moments with $l=0$. We found that the Chapman-Enskog approximation better agrees with the exact solution, especially for small masses and higher-order moments, compared to the 14-moment approximation (see fig.~\ref{fig:enter-label}). Furthermore, we found that at late times, the first-order gradient expansion does not sufficiently explain the bulk viscous correction. Moreover, our analysis shows that there is a consistent approach to the forward attractor for all moments and a semi-universal behavior of the early-time dynamics for $l = 0$ moments for phenomenologically relevant initialization times.  This semi-universality translates into a slight uncertainty of the attractor in non-conformal systems, but still supports the utility of attractors in heavy-ion collision models.

\bibliography{attractorQM}

\end{document}